\documentstyle[12pt]{article}

\def\be{\begin{equation}}
\def\ee{\end{equation}}
\def\ben{$$}
\def\een{$$}
\def\ba{\begin{array}{c}}
\def\ea{\end{array}}
\def\p{\partial}
\begin{document}

\titlepage
\vspace*{2cm}

\begin{center}{\large \bf
ELEMENTARY DOUBLETS OF BOUND STATES OF THE RADIAL DIRAC EQUATION
 }\end{center}

\vspace{5mm}

\begin{center}
Miloslav Znojil
\vspace{3mm}

\'{U}stav jadern\'e fyziky AV \v{C}R, 250 68 \v{R}e\v{z},
Czech Republic\\

\end{center}

\vspace{5mm}

\section*{Abstract}

For non-relativistic Schr\"{o}dinger equations the lowering of
their degree by substitution $\Psi(r) \to F(r) =\Psi'(r) /\Psi
(r)$ is known to facilitate our understanding and use of their
(incomplete, so called quasi-exact) solvability.  We show that
and how the radial Dirac relativistic equation may quasi-exactly
be solved in similar spirit.

\vspace{5mm}

PACS 03.65.Ge

\newpage

\section{Introduction}

The two-component radial Dirac equation with energy $E$, mass
$M$ and centrifugal term $U(r)=\kappa/r$ \cite{Bjorken} reads
\be
\left[
 \begin{array}{cc} \partial_r-U(r)&M+W(r)-E-V(r)\\ M+W(r)+E+V(r)
&\partial_r+U(r)
\ea \right] \left[
\ba f(r)\\g(r) \ea \right] =0 .
\label{Dir}
\ee
In a way similar to non-relativistic Schr\"{o}dinger equation it
proves exactly solvable for the electrostatic field of hydrogen
atom $V(r) = \alpha/r$ accompanied, if necessary, by the
auxiliary or external Lorentz scalar force $W(r) = \beta/r$ and
by the possible central magnetic monopole charge $Q$ in $\kappa
=\pm\sqrt{(\ell +1)(\ell+1+2|Q|)}$ where $\ell = 0, 1,\ldots$.
Marginally, let us note that $\kappa=\ell+1=0$ is also admitted
whenever $\beta^2 >\alpha^2$ and $Q\neq 0$ (cf. ref.
\cite{Torres} for details).

Recently, Brihaye and Kosinski \cite{Brihaye} conjectured that a
formal parallel between relativistic and non-relativistic
quantum mechanics may be extended to many other models.
Explicitly, they have demonstrated that the perturbation of
hydrogen atom by the linear relativistic force $W(r) \sim
\omega\,r$ not only resembles its non-relativistic Coulomb plus
linear plus quadratic analogue but also shares the {\em
incomplete}, so called quasi-exact (QE) solvability with it.  At
certain exceptional energies and couplings, elementary bound
states were obtained by non-numerical means in a way which
complements the 25 years old non-relativistic result by A.
Hautot
\cite{Hautot}.

In our present letter we intend to proceed one step further.
Having in mind a deep non-relativistic connection between QE
solvability and Riccati-Schr\"{o}dinger equations
\cite{Ushveridze}, we shall formulate a parallel relativistic
``order-lowering" idea and implement it in the technically
slightly more difficult context of relativistic eq. (\ref{Dir}).
This will enable us to show that, in particular, the
algebraic-equation approach of Brihaye and Kosinski just picks
up a very specific portion of a much larger class of all the QE
solvable Dirac equations.

\section{New QE solutions: Explicit method}

Once we represent a wavefunction in non-relativistic quantum
mechanics as an integral $\Psi_0(r) =\exp
{\int^r_{r_{ini}}F(\xi) d\xi} $ we get $\Psi''_0(r,\ell
)/\Psi_0(r,\ell) = [F(r)]^2+\p_r F(r) $.  This converts the
radial differential Schr\"{o}dinger equation to an equivalent
first-order form $V(r)= E_0 - {\ell(\ell+1) / r^2}  +
[F(r)]^2+\p_r F(r) $.  Such a Riccati-type re-arranged equation
is nonlinear but may be re-interpreted as an explicit closed
definition of a partially solvable potential in terms of any of
its ``tentative" wavefunctions.

The latter point of view plays an important role in the
understanding of non-relativistic QE systems \cite{Ushveridze}.
{\it Mutatis mutandis}, the relativistic pair of equations
(\ref{Dir}) may also define the QE solvable pairs of potentials
$W_{}(r)$ and $V_{}(r)$.  Indeed, assuming that the latter
forces are responsible for the existence of any particular
elementary wavefunction with components $f_0(r)$ and $g_0(r)$ at
a particular mass $M_0$ and energy $E_0$ we may write
\ben
W_{}(r)=-M_0-{1 \over 2}\left[ {1\over g_0}\left(\p_r-U\right)f_0 +
{1\over f_0}\left(\p_r+U\right)g_0
\right],
\een
\be
V_{}(r)=-E_0+{1 \over 2}\left[ {1\over g_0}\left(\p_r-U\right)f_0 -
{1\over f_0}\left(\p_r+U\right)g_0
\right].
\label{star}
\ee
On the basis of experience with non-relativistic models it is
not too surprising that the mere correct threshold and
asymptotic behaviour in tentative $f_0(r) = p\,r^\mu \exp
(-\lambda \,r)$ and $g_0(r) = q\,r^\mu \exp (-\lambda \,r)$ with
normalization constants $p$ and $q$ already leads to the exactly
and, incidentally, completely solvable model as mentioned above
(cf. also \cite{Brihaye}).  The next tentative elementary choice
could mimick an unphysical singularity at negative $r_u=-1/h<0$,
\be
f_0(r) = p\,r^\mu (1+h\,r)\,\exp (-\lambda \,r), \ \ \ \ \ \
g_0(r) = q\,r^\mu (1+h\,r)\, \exp (-\lambda \,r).
\ee
In terms of the same parameters $\varepsilon = \pm 1$ and $t \in
(-\infty,\infty)$ in the input ratio of norms $p/q
=\varepsilon\,\exp t$ we get the same formula for energy $E =
-\varepsilon\,\lambda\,\sinh t$ and mass $M
=\varepsilon\,\lambda\,\cosh t$ as above.  Also both the
Coulombic couplings remain the same,
\be
\left( \begin{array}{c}
\beta\\
\alpha\ea \right)
=
\left( \begin{array}{cc}
-\cosh t &\sinh t \\
\sinh t&- \cosh t
\ea \right)
\left(
 \begin{array}{c}
\varepsilon\,\mu\\
\varepsilon\,\kappa
\ea \right).
\label{erzeta}
\ee
The only change emerges as a screening which enters the new and,
by definition, QE solvable potentials
\be
V(r) = {\alpha \over r} +  {\alpha_s \over 1+h\,r},\ \ \ \ W(r)
= {\beta \over r} +  {\beta_s \over 1+h\,r},\ \ \ \
\alpha_s=\varepsilon\,h\,\sinh t, \ \ \ \
\beta_s=-\varepsilon\,h\,\cosh t.
\label{newm}
\ee
Further states in such a model may be sought and studied, by the
explicit algebraic method of ref. \cite{Brihaye}, in full
analogy with semi-relativistic and non-relativistic QE solvable
screened Coulomb potentials \cite{94a12083}.

\section{QE solutions: Implicit method}

Our new QE model (\ref{newm}) looks particularly simple after
transition to the elementary integral representation of
wavefunctions
\be
f(r) = e^{\int^r_{r_{ini}}F(\xi) d\xi}, \ \ \ \ \ \ \ g(r) =
e^{\int^r_{r_{ini}}G(\xi) d\xi}.
\label{libera}
\ee
In non-relativistic setting, similar re-arrangement proved
useful in computations \cite{Fernandez} as well as in the so
called supersymmetric transformations of Hamiltonians
\cite{Cooper}.  Also here, our postulate (\ref{libera}) will
lead to simplifications.  Thus, with abbreviations $ Y(r) =
[F(r) + G(r) ]/ 2$ and $ Z(r) = [F(r) - G(r)]/ 2$, one of the
components of Dirac equation lowers its order and becomes purely
algebraic,
\be
[E+V(r)]^2 + [Y(r)]^2 = [M+W(r)]^2 + [U(r)-Z(r)]^2.
\label{eqjedna}
\ee
This enables us to parametrize, say,
\be
E+V(r) = R(r) \cos A(r), \ \ \ \ \ \ \ M+W(r) = R(r) \cos B(r),
\label{potiksa}
\ee
\be
 Y(r) = R(r) \sin A(r), \ \ \ \ \ \ \ \ \
\ U(r) - Z(r) = R(r) \sin B(r).
\label{tri}
\ee
Such a transformation simplifies also the remaining Dirac
equation
\be
 M+W(r) - E-V(r)+[Y(r)+Z(r)-U(r)]\,\exp 2 \int^r_{r_{ini}}
Z(\xi) d \xi = 0
\label{eqdva}
\ee
which, in terms of an abbreviation $C(r)=-[A(r) +B(r)]/ 2$,
reads $ {\rm tan}\, C(r) =\exp 2 \int^r_{r_{ini}} Z(\xi) d \xi $
or, in differential form, $Z(r) = {\p_r C(r) / \sin 2C(r)}$.
After an insertion of such a definition of $Z(r)$ in the second
item of eq. (\ref{tri}) we may eliminate the function $R(r)$ and
are left with a set of the closed and compact simultaneous
definitions of both the wavefunctions {\em and} potentials in
terms of an arbitrary initial choice of the auxiliary but
practically unrestricted pair of functions $A(r)$ and $B(r)$.

\section{Re-parametrization}

Our above construction of a QE solvable system exhibits still a
certain similarity to its non-relativistic Riccati-like version.
Unfortunately, the parallel is incomplete.  In particular, we
cannot derive potentials immediately from the wavefunctions
since {\it both} of them enter our formulae together with their
derivatives.  At the same time, our implicit QE-type solution of
Dirac eq. (\ref{Dir}) still has to be made compatible with some
overall physical requirements.  Also its clearer physical
interpretation is needed.  For this purpose, let us make a
further step.  Recalling the standard Pauli matrices
\be
\sigma_x=
\left( \begin{array}{cc} 0&1\\1&0
\ea \right), \ \ \ \ \ \ \ \
i\sigma_y=
\left(
 \begin{array}{cc} 0&1\\-1&0
\ea \right), \ \ \ \ \ \ \ \
\sigma_z=
\left(
 \begin{array}{cc} 1&0\\0&-1
\ea \right)
\ee
let us re-write eq. (\ref{Dir}) as a manifestly real
two-component problem
\be
\left\{ I\,\p_r +[M+W(r)]\,\sigma_x
-i[E+V(r)]\,\sigma_y-U(r)\,\sigma_z
\right\} \, \psi(r)=0
\ee
and pre-multiply it by the transposed two-component real spinors
$\sigma_x\psi(r)$ and $i\sigma_y\psi(r)$ from the left.  The
resulting pair of relations
\be
\left[
 \begin{array}{c} Z(r)-U(r)\\-Y(r)
\ea \right]=
\left[
 \begin{array}{cc} \cosh \Xi(r) &\sinh \Xi(r)\\
\sinh \Xi(r)& \cosh \Xi(r)
\ea \right]
\left[
 \begin{array}{c} E+V(r)\\ M+W(r)
\ea \right]
\ee
with $\Xi(r) = 2\int^r_{r_{ini}}Z(\xi)d\xi$ represents another,
integral representation of our original Dirac bound-state
problem.  In it, the two-by-two matrix is easily invertible,
\be
\left[
 \begin{array}{c} E+V(r)\\ M+W(r)
\ea \right]
=\left[ \begin{array}{cc} \cosh \Xi(r) &-\sinh \Xi(r)\\ -\sinh
\Xi(r)& \cosh \Xi(r)
\ea \right]
\left[
 \begin{array}{c} Z(r)-U(r)\\-Y(r)
\ea \right].
\label{verze}
\ee
This induces the simplified {\it ansatz}
\be
E+V(r) = S(r) \sinh T(r), \ \ \ \ \ \ \ M+W(r) = S(r) \cosh
T(r),
\label{zatik}
\ee
\be
 Z(r) = U(r) + S(r) \sinh[ T(r)+\Xi(r)], \ \ \ \ \ \ \ \ \ Y(r)
=- S(r) \cosh[T(r)+\Xi(r)]
\label{zatri}
\ee
which, in particular, parametrizes all the QE solutions by the
independent input functions $S(r)$ and $T(r)$.  As already
mentioned above, they must only be subject to the appropriate
physical boundary conditions.  We may conclude that the {\em
implicit} relativistic implementation of the idea of QE
solvability is as straightforward as its {\em explicit}
nonrelativistic predecessor.

\section{Relativistic QE doublets}

After any change of our above ``parametrization" point of view,
technical complications may re-emerge immediately.  Even for the
elementary and popular Coulomb + polynomial form of forces as
suggested for further study of the QE solvability in ref.
\cite{Brihaye} we immediately imagine that a seemingly trivial
guarantee of their compatibility with our parameterizations
leads in fact to a quite difficult algebraic problem.

Another, mathematically easier {\em and} practically more
important question is the possible existence and/or feasibility
of constructions of the elementary QE multiplets.  Indeed, in
principle, after any choice of a QE wavefunction the resulting
partially solvable potential may still remain compatible with
{\em another} elementary bound state.  Even in non-relativistic
case such a physically useful requirement is mathematically
non-trivial \cite{Leach}. There, it may still be characterized
by the comparatively transparent condition
\be
\p_r[F_1(r) - F_2(r)] + [F_1(r) - F_2(r)][(F_1(r)+F_2(r)] =
E_2-E_1.
\label{hujer}
\ee
For the {\em same} QE potential (which was eliminated) this
equation defines the superposition $F_s(r)=[ F_1(r) + F_2(r)]/2$
in terms of the differences $F_d(r)=[F_1(r) - F_2(r)]/2$ and
$\delta = E_2-E_1$ \cite{Ushveridze}.  This means that there
exist very many nonrelativistic doublet partners $F_1=F_s+F_d$
and $F_2 =F_s-F_d$ which are ``numbered" by the choice of the
virtually unconstrained functions $F_d(r)$.

In relativistic case, we have to proceed in similar vain.  The
elimination of the two energy-independent QE solvable potentials
$V(r)$ and $W(r)$ from the two versions of eq.  (\ref{verze})
will describe the difference between the two right hand sides as
an $r-$independent spinor proportional to $\delta$.  We get a
relativistic differential-equation counterpart to eq.
(\ref{hujer}) in terms of the auxiliary integrals
$\alpha=\alpha(r) =\int^r(Z_2-Z_1)$ and $\beta=\beta(r)
=\int^r(Z_2+Z_1)$ and their derivatives
$Z_2(r)=\p_r\alpha/2+\p_r\beta/2$ and $Z_1(r)= -\p_r\alpha
/2+\p_r\beta/2$,
\be
\left( \begin{array}{c}
Z_2
-U\\ -Y_2
\ea \right)
=
\delta\,
\left[ \begin{array}{c}
\cosh(\alpha+\beta)\\
\sinh(\alpha+\beta)
\ea \right]
+
\left( \begin{array}{cc}
\cosh 2\alpha &\sinh 2\alpha \\
\sinh 2\alpha& \cosh 2\alpha
\ea \right)
\left(
 \begin{array}{c} Z_1-U\\ -Y_1
\ea \right).
\label{konverze}
\ee
This formula may be read as an algebraic linear set of
definitions of the two sums of exponents $Y_1(r)$ and $Y_2(r)$.
The symmetry of our new equation with respect to the
simultaneous double reflection $\beta(r)\leftrightarrow
-\beta(r)$, $U(r)\leftrightarrow -U(r)$ and permutation
$Y_1(r)\leftrightarrow Y_2(r)$ is one of the reasons why its
closed solution is still unexpectedly compact,
\be
Y_1(r) + Y_2(r) = \delta\, {\cosh \beta(r) \over \sinh
\alpha(r)} -  \alpha'(r)\, {\cosh \alpha(r) \over \sinh
\alpha(r)} ,
\ee
\be
Y_1(r) - Y_2(r) = \delta \,{\sinh \beta(r) \over \cosh
\alpha(r)} +
[\beta'(r)-2U(r)]\, {\sinh \alpha(r) \over \cosh
\alpha(r)} .
\ee
We may summarize that our ``parameters" $\alpha(r)$ and
$\beta(r)$ remain practically arbitrary functions.  In the
relativistic Dirac case there exist infinitely many QE
bound-state doublets as well.  Thus, the well known {\em
functional} freedom of the implicit doublet solutions of the
non-relativistic Schr\"{o}dinger QE equations is shared by our
present Dirac relativistic QE construction.

\section*{ Acknowledgments}

Partially supported by the {\em Grantov\'{a} agentura AV \v{C}R,
Praha} (grant Nr. A 104 8602) and by the {\em Grantov\'{a} agentura
\v{C}R, Praha} (grant Nr. 202/96/0218).


\newpage

\begin{thebibliography}{00}

\bibitem{Bjorken}
H. A. Bethe and E. E. Salpeter, Quantum Mechanics of One- and
Two-Electron Atoms (Academic Press, New York, 1957);

J. D. Bjorken and S. D. Drell, Relativistic Quantum Mechanics
(McGraw-Hill, New York, 1964), p. 52;

A. E. S. Green, in Antinucleon- and Nucleon-Nucleus
Interactions, ed. by G. E. Walker et al (Plenum, New York,
1985), p. 143;

C. J. Horowitz, in Relativistic Dynamics and Quark-Nuclear
Physics, ed. by M. B. Johnson and A. Picklesimer (Wiley, New
York, 1986), p. 221;

C. Semay, R. Ceuleneer and B. Silvestre-Brac, J. Math. Phys. 34,
2215 (1993);

V. Villalba, J. Math. Phys. 36, 3332 (1995).

\bibitem{Torres}
G. Torres del Castillo and L. Cortes-Curantli, J. Math. Phys.
38, 2996 (1997).

\bibitem{Brihaye}
Y. Brihaye and P. Kosinski, Mod. Phys. Lett. A 13, 1445 (1998).

\bibitem{Hautot}
A. Hautot, Phys. Lett. A 38, 305 (1972).

\bibitem{Ushveridze}
A. G. Ushveridze, Quasi-exactly solvable models in quantum
mechanics (IOPP, Bristol, 1994).

\bibitem{94a12083}
M. Znojil, Phys. Lett. A 94, 120 (1983) and J. Phys. A: Math.
Gen. 29, 6443 (1996).

\bibitem{Fernandez}
F. M. Fern\'{a}ndez, G. I. Frydman and E. A. Castro, J. Phys. A:
Math. Gen. 22, 641 (1989);

F. M. Fern\'{a}ndez, Q. Ma and R. H. Tipping, Phys. Rev. A 39,
1605 (1989) and A 40, 6149 (1989);

V. C. Aguilera-Navarro, F. M. Fern\'{a}ndez, R. Guardiola and J.
Ros, J. Phys. A: Math. Gen. 25, 6379 (1992);

F. M. Fern\'{a}ndez and R. Guardiola, J. Phys. A: Math. Gen. 26,
7169 (1993);

F. M. Fern\'{a}ndez, R. Guardiola and M. Znojil, Phys. Rev. A
48, 4170 (1993);

M. Znojil, J. Phys. A: Math. Gen. 28, 6265 (1995).

\bibitem{Cooper}
F. Cooper, A. Khare and U. Sukhatme, Phys. Rep. 251, 267 (1995).

\bibitem{Leach}
M. Znojil and P. G. L. Leach, J. Math. Phys. 33, 2785 (1992);

M. Znojil and R. Roychoudhury, Czech. J. Phys. 48, 1 (1998).

\end{thebibliography}
\end{document}